\documentclass[11pt,psfig]{article}
\usepackage{amsmath}
\usepackage{setspace}
\usepackage{mathrsfs}
\usepackage{amsfonts}
\usepackage{graphics}
\usepackage{graphicx}
\usepackage{latexsym}
\usepackage{amsthm}
\usepackage{cite}
\usepackage{amssymb}
\usecounter{enumi}
\newtheorem{Theorem}{Theorem}

\newtheorem{Lemma}{Lemma}
\newtheorem{Definition}{Definition}
\newtheorem{Corollary}{Corollary}

\numberwithin{equation}{section}
\numberwithin{Theorem}{section}
\numberwithin{Lemma}{section}
\numberwithin{Definition}{section}
\numberwithin{Corollary}{section}

\def\be{\begin{eqnarray}} \def\ee{\end{eqnarray}} 
  \def\({\left(} \def\){\right)}
\def\bc{\begin{center}} 
\def\ec{\end{center}}  

\def\bey{\begin{eqnarray*}}\def\eey{\end{eqnarray*}}

\begin{document}
\title{The Quasi-Integrability of a Generalized Camassa-Holm Equation}
\author{ Mingyue Guo$^{a}$, Zhenhua Shi$^{a,b}$ \footnote{Corresponding author, E-mail address: zhenhuashi@nwu.edu.cn}
\vspace{4mm}\\
$^{a}$\small School of Mathematics, Northwest University, Xi'an 710069, China\\
$^{b}$\small Center for Nonlinear Studies, Northwest University, Xi'an 710069, China}
\date{}
\maketitle
 \bc
\begin{minipage}{120mm}
{\bf Abstract}\\
This paper examines a generalization of the Camassa-Holm equation from the perspective of integrability. Using the framework developed by Dubrovin on bi-Hamiltonian deformations and the general theory of quasi-integrability, we demonstrate that a unique bi-Hamiltonian structure is possible for this generalized equation only when it reduces to the original CH equation.

\end{minipage}
\ec

\section{Introduction}

\indent \indent The Camassa-Holm (CH) equation, expressed as:
\begin{equation}
u_t-u_{xxt}=3uu_x-2u_xu_{xx}-uu_{xxx}
\end{equation}
for $u(t,x)$,  has attracted significant attention in the last three decades due to its relevance in shallow water wave theory. In 1993, it was shown \cite{ref1, ref2} that this equation arises from the theory of shallow water waves, derived via an asymptotic expansion of Euler's equations for inviscid fluid flow. Since then, research efforts have focused extensively on exploring its mathematical and physical properties, alongside generalizations that maintain some of its key features. A notable family of generalized CH equations is formulated as:
\begin{equation}
u_t-u_{xxt}=\frac{1}{2}(p+1)(p+2)u^pu_{x}-\frac{1}{2}p(p-1)u^{p-2}u_x^3-2pu^{p-1}u_xu_{xx}-u^pu_{xxx},
\end{equation}
where $p \in \mathbb{Z} \setminus \{ 0 \}$ is a parameter representing nonlinearity. This family of equations was inspired by one of the Hamiltonian structures of the CH equation and draws analogies with the generalized KdV family \cite{ref3, ref4}
\begin{equation}
u_t-u^p u_x-u_{xxx}=0,\quad p\neq 0.
\end{equation}
This generalized CH equation  has been shown to admit single peakon solutions but does not allow for multi-peakon solutions \cite{ref5}.

Integrability is a relatively rare property for partial differential equations, and several definitions (related but not equivalent) exist \cite{ref6, ref7, ref8, ref9}. For this paper, we define integrability as follows:
\begin{equation}
u_t=-B_1\frac{\delta\mathcal{H}_1}{\delta u}=-B_2\frac{\delta\mathcal{H}_2}{\delta u},
\end{equation}
where $B_1$ and $B_2$ are the Hamiltonian operators satisfying the Jacobi identity, $\mathcal{H}_1$ and $\mathcal{H}_2$ are the Hamiltonian functions such that any linear combination of Hamiltonian operators remains a Hamiltonian operator.

It is well-established that equation (1.2) is integrable when $p=1$, which reduces to the CH equation. However, for arbitrary choices of $p$, we have not found any explicit results indicating integrability beyond the case of the CH equation. In this paper, we apply a perturbative method due to Dubrovin \cite{ref10,ref11} to investigate the quasi-integrability of equation (1.2), a weaker notion of integrability introduced by da Silva PL and Freire IL \cite{ref12}.

\begin{Definition}
An equation is said to be quasi-integrable if, eventually under a change of variables, it satisfies Lemmas 2.1 and 2.2.
\end{Definition}

Lemmas 2.1 and 2.2, presented in Section 2, essentially demonstrate the existence of an infinite hierarchy of approximate symmetries \cite{ref13} and uniqueness of a bi-Hamiltonian deformation of hyperbolic equations. We have the following result about the quasi-integrability of Equation (1.2):
\begin{Theorem}
Equation (1.2) is quasi-integrable if and only if $p=1$. In particular, its perturbed Hamiltonian is
\begin{equation*}
\mathcal{H}=\int\left(-\frac{1}{2}v^3+\epsilon^2 vv_x^2-\epsilon^4 vv_{xx}^2\right)dx,\quad v=(1-\epsilon^2\partial_x^2)^{1/2}u.
\end{equation*}
Moreover, the equation (1.2) admits a unique bi-Hamiltonian structure given by
\begin{equation*}
  \begin{split}
    \{v(x),v(y)\}_1&=\delta^{\prime}(x-y),\\
    \{v(x),v(y)\}_2&=v\delta^{\prime}(x-y)+\frac{1}{2}v_x\delta(x-y)
+\epsilon^2\left[v\delta^{\prime\prime\prime}(x-y)+\frac{3}{2}v_x\delta^{\prime\prime}(x-y)+\frac{1}{2}v_{xx}\delta^{\prime}(x-y)\right]
    \\&+\epsilon^4\left[v\delta^{\prime\prime\prime\prime\prime}(x-y)+\frac{5}{2}v_x\delta^{\prime\prime\prime\prime}(x-y)+3v_{xx}\delta^{\prime\prime\prime}(x-y)
    +2v_{xxx}\delta^{\prime\prime}(x-y)+\frac{1}{2}v_{xxxx}\delta^{\prime}(x-y)\right].
   \end{split}
\end{equation*}
\end{Theorem}

\begin{Corollary}
Equation (1.2) is quasi-integrable if and only if it is equivalent to the CH equation.
\end{Corollary}

\section{Preliminaries}
\indent \indent To initiate our analysis, we recall key results from \cite{ref10}. We consider Hamiltonian perturbations of the hyperbolic equation:
\begin{equation}
v_t+a(v)v_x=0,
\end{equation}
which is known to exhibit infinitely many symmetries and conservation laws (e.g., see Refs. 14, 15 and references therein). This equation can be written in Hamiltonian form as follows:
\begin{equation}
v_t+\{v(x),\mathcal{H}_0\}=0,\quad \mathcal{H}_0=\int f(v)dx,\quad f^{\prime\prime}(v)=a(v),
\end{equation}
where $\mathcal{H}_0$ is the unperturbed Hamiltonian, while $\{\cdot,\cdot\}$ denotes a poisson bracket. In the remainder of this paper, $\{\cdot,\cdot\}_1$ and $\{\cdot,\cdot\}_2$ represent two distinct Poisson brackets, adhering to Dubrovin's notation. We will adhere closely to Dubrovin's notation throughout the remainder of the discussion.
\begin{Lemma}
For any $f=f(v)$ the Hamiltonian flow
\begin{equation}
\begin{split}
  &v_t+\partial_x\frac{\delta\mathcal{H}_f}{\delta v}=0,
\\&\mathcal{H}_f=\int h_f(v)dx,
\\&h_f=f-\epsilon^2 \frac{c}{24}f^{\prime\prime\prime}v^2_x+\epsilon^4\left[\left(rf^{\prime\prime\prime}+\frac{c}{480}f^{(4)}\right)v^2_{xx}\right.
\\&\left.-\left(\frac{cc^{\prime\prime}}{1152}f^{(4)}+\frac{cc^{\prime}}{1152}f^{(5)}+\frac{c^2}{3456}f^{(6)}
+\frac{r^{\prime}}{6}f^{(4)}+\frac{r}{6}f^{(5)}-sf^{\prime\prime\prime}\right)v^4_x\right],
\end{split}
\end{equation}
where $c=c(v)$ and $r=r(v)$, is a symmetry, module $\mathcal{O}(\epsilon^6)$ of (2.2). Moreover, the Hamiltonian $\mathcal{H}_f$ commute pairwise, in the sense that $\{\mathcal{H}_f,\mathcal{H}_g\}=\mathcal{O}(\epsilon^6)$ for arbitrary two functions $f(v)$ and $g(v)$.
\end{Lemma}

Not all perturbations are bi-Hamiltonian, but we have a very interesting and valuable result.
\begin{Lemma}
If $c(v)\neq0$, the commuting Hamiltonians (2.3) admit a unique bi-Hamiltonian structure obtained by a deformation of
\begin{equation*}
\{v(x),v(y)\}_1=\delta^{\prime}(x-y),\quad \{v(x),v(y)\}_2=q(v(x))\delta^{\prime}(x-y)+\frac{1}{2}q^{\prime}(v)v_x\delta(x-y)
\end{equation*}
with the functions $r(v)$, $q(v)$, $c(v)$ and $s(v)$ satisfying the relation
\begin{equation}
r(v)=\frac{c(v)^2}{960}\left(\frac{5c^{\prime}(v)}{c(v)}-\frac{q^{\prime\prime}(v)}{q^{\prime}(v)}\right),\quad s(v)=0.
\end{equation}
\end{Lemma}

\section{Integrability Results}
\indent \indent To prove Theorem 1.1, we apply the techniques introduced by Dubrovin. The crux of determining whether equation (1.2) satisfies the conditions of Definition 1.1 lies in treating its higher-order derivative terms as perturbations of a hyperbolic equation of the form (2.1). For this, we introduce the change of variables:
\begin{equation*}
(t,x,u)\mapsto(\epsilon t,\epsilon x, u),\quad 0\neq\epsilon\ll 1,
\end{equation*}
into the variables of equation (1.2). This rescaling transforms equation (1.2) into the following:
\begin{equation}
u_t-\epsilon^2u_{xxt}-\frac{1}{2}(p+1)(p+2)u^pu_{x}+\epsilon^2\left(\frac{1}{2}p(p-1)u^{p-2}u_x^3-2pu^{p-1}u_xu_{xx}-u^pu_{xxx}\right)=0.
\end{equation}

\noindent \textbf{Proof of Theorem 1.1.} Next, observe that Lemmas 2.1 and 2.2 specifically apply to evolution equations, while equation (3.1) is not initially expressed in this form. To address this, we rewrite $u-\epsilon^2\partial^2_x u=(1-\epsilon^2\partial^2_x)u$, and since this expression represents a perturbation of the identity operator, it is invertible. For $s\in\mathbb{R}$, the operator can be expanded as
\begin{equation}
(1-\epsilon^2\partial^2_x)^{s/2}=1-\frac{s}{2}\epsilon^2\partial^2_x+\frac{s(s-2)}{8}\epsilon^4\partial^4_x+\mathcal{O}(\epsilon^6)\approx 1-\frac{s}{2}\epsilon^2\partial^2_x+\frac{s(s-2)}{8}\epsilon^4\partial^4_x.
\end{equation}

It is worth mentioning that this operator commutes with $\partial_t$. We can now transform the non-evolution equation (3.1) into an evolution equation by following these steps:
\begin{itemize}
  \item We begin by writing  $m=u-\epsilon^2\partial^2_x u$ and define $v=(1-\epsilon^2\partial^2_x)^{-1/2}m$, which implies that $u=(1-\epsilon^2\partial^2_x)^{-1/2}v$;
  \item Next, we rewrite Equation (3.1) as follows:
  \begin{equation}
  \begin{split}
  (1-\epsilon^2\partial^2_x)^{-1/2}m_t  & =(1-\epsilon^2\partial^2_x)^{-1/2}\left(\frac{1}{2}(p+1)(p+2)u^pu_{x}\right)
   \\ & -\epsilon^2(1-\epsilon^2\partial^2_x)^{-1/2}\left(\frac{1}{2}p(p-1)u^{p-2}u_x^3-2pu^{p-1}u_xu_{xx}-u^pu_{xxx}\right);
  \end{split}
  \end{equation}
  \item Then, we use the expansion (3.2) with $s=-1$ to obtain the following series expansions for the terms involving $u$:
  \begin{equation*}
  \begin{split}
   &u = v+\frac{\epsilon^2}{2}v_{xx}+\frac{3}{8}\epsilon^4 v_{xxxx}+\mathcal{O}(\epsilon^6),
  \\& u_x=v_x+\frac{\epsilon^2}{2}v_{xxx}+\frac{3}{8}\epsilon^4 v_{xxxxx}+\mathcal{O}(\epsilon^6),
  \\& u_{xx}=v_{xx}+\frac{\epsilon^2}{2}v_{xxxx}+\frac{3}{8}\epsilon^4 v_{xxxxx}+\mathcal{O}(\epsilon^6),
  \\& u_{xxx}=v_{xxx}+\frac{\epsilon^2}{2}v_{xxxxx}+\frac{3}{8}\epsilon^4 v_{xxxxxxx}+\mathcal{O}(\epsilon^6);
  \end{split}
  \end{equation*}
  \item Finally, we use the relation $ (1-\epsilon^2\partial^2_x)^{-1/2}m_t=v_t$, and substitute the above expressions and the expansion (3.2) with $s=-1$ into Equation (3.3).
\end{itemize}

The final result of the above procedure is the equation
\begin{equation}
v_t = \text{I} + \epsilon^2 \left( \text{II} + \frac{1}{2} \partial^2_x \text{I} \right)
+ \epsilon^4 \left( \text{III} + \frac{1}{2} \partial^2_x \text{II} + \frac{3}{8} \partial^4_x \text{I} \right) + \mathcal{O}(\epsilon^6),
\end{equation}
whose term are
\begin{equation}
  \begin{split}
  & \text{I} =\frac{1}{2}(p+1)(p+2)v^p v_x,
\\& \text{II}=\frac{1}{2}p(p+3)v^p v_{xxx}+p^2(p+3)v^{p-1}v_x v_{xx}+\frac{1}{4}p^2(p-1)(p+3)v^{p-2}v_x^3,
\\&\text{III}=\frac{1}{2}p(p+3)v^p v_{xxxxx}+\frac{5}{2}p^2(p+3)v^{p-1}v_{xx}v_{xxx}+\frac{3}{2}p^2(p+3)v^{p-1}v_x v_{xxxx}
\\&+\frac{13}{4}p(p-1)\left(p^2+3p+\frac{4}{13}\right)v^{p-2}v_xv_{xx}^2+\frac{9}{4}p(p-1)\left(p^2+3p+\frac{2}{9}\right)v^{p-2}v_x^2v_{xxx}
\\&+2p(p-1)(p-2)\left(p^2+3p+\frac{1}{2}\right)v^{p-3}v_x^3 v_{xx}+\frac{3}{16}p(p-1)(p-2)(p-3)\left(p^2+3p+\frac{2}{3}\right)v^{p-4}v_x^5.
  \end{split}
\end{equation}
From now on, we neglect terms $\mathcal{O}(\epsilon^6)$. Then Hamiltonian in (2.3) takes the form
\begin{equation}
\mathcal{H}_f=\mathcal{H}_0+\epsilon^2 \mathcal{H}_1+\epsilon^4 \mathcal{H}_2,
\end{equation}
where
\begin{equation}
\begin{split}
  &\mathcal{H}_0=\int f(v)dx,\quad \mathcal{H}_1=\int \frac{c}{24}f^{\prime\prime\prime}v_x^2 dx,
\\& \mathcal{H}_2=\int\left[\left(rf^{\prime\prime\prime}+\frac{c}{480}f^{(4)}\right)v^2_{xx}-\left(\frac{cc^{\prime\prime}}{1152}f^{(4)}
+\frac{cc^{\prime}}{1152}f^{(5)}+\frac{c^2}{3456}f^{(6)}
+\frac{r^{\prime}}{6}f^{(4)}+\frac{r}{6}f^{(5)}-sf^{\prime\prime\prime}\right)v^4_x\right]dx.
\end{split}
\end{equation}

Our goal here is to investigate whether Equation (3.4) can be expressed in the form of Equation (2.3). Assuming this hypothesis, we proceed by considering the coefficients of $1$, $\mathcal{O}(\epsilon^2)$ and $\mathcal{O}(\epsilon^4)$ in the expansions given by (3.5), (3.7), and the first equation in (2.3). This leads to the following system of equations for the respective terms:
\begin{equation}
 \text{I}=-\partial_x\frac{\delta\mathcal{H}_0}{\delta v},
\end{equation}
\begin{equation}
\text{II} + \frac{1}{2} \partial^2_x \text{I}=-\partial_x\frac{\delta\mathcal{H}_1}{\delta v},
\end{equation}
\begin{equation}
 \text{III} + \frac{1}{2} \partial^2_x \text{II} + \frac{3}{8} \partial^4_x \text{I}=-\partial_x\frac{\delta\mathcal{H}_2}{\delta v}.
\end{equation}

We now use Equation (3.8) to determine whether and under what conditions Equation (1.2) satisfies Lemma 2.1. From this point onward, we assume all constants of integration are zero. Based on Equations (3.5) and (3.8), we conclude that
\begin{equation}
f(v)=-\frac{1}{2}v^{p+2}.
\end{equation}
Substituting Equation (3.11) into $\mathcal{H}_1$ in (3.7), and using Equations (3.5) and (3.9), we obtain
\begin{equation}
\frac{1}{4}p^2(p+3)v^{p-1}v_x^2+\frac{1}{2}p(p+3)v^{p}v_{xx}=\frac{1}{48}p(p+1)(p+2)\left[c^{\prime}v^{p-1}v_x^2+(p-1)cv^{p-2}v_x^2+2cv^{p-1}v_{xx}\right].
\end{equation}
Matching the coefficients of $v_x^2$ and $v_{xx}$  from both sides, we arrive at the following system of equations:
\begin{equation}
\frac{1}{4}p^2(p+3)v=\frac{1}{48}p(p+1)(p+2)\left[c^{\prime}v^{p-1}+(p-1)c\right],
\end{equation}
and
\begin{equation}
\frac{1}{2}p(p+3)v=\frac{1}{24}p(p+1)(p+2)c,
\end{equation}
which gives
\begin{equation}
c(v)=\frac{12(p+3)}{p^2+3p+2}v.
\end{equation}
Following the same steps as before, we substitute Equations (3.11) and (3.15) into $\mathcal{H}_2$ in (3.7) and use Equation (3.5). By matching the coefficients of $v_{xxxx}$ and $v_x^2 v_{xx}$ in result of (3.10), we get
\begin{equation}
r(v)=\frac{p+3}{2(p^2+3p+2)}v-\frac{p^2+2p+3}{40(p^2+3p+2)},
\end{equation}
and
\begin{equation}
s(v)=-\frac{p^5+6p^4+6p^3-9p^2-4p}{6p(p^2+3p+2)^2v}-\frac{p^5-p^4-7p^3+13p^2-6p}{240p(p^2+3p+2)v^2}.
\end{equation}
Consider that Equation (1.2) satisfies Lemma 2.2 as well,  this would occur only when $p=1$, which implies that $r(v)=\frac{1}{3}v$ and $q(v)=v$, where $a$ and $b$ are two arbitrary constants, with $a\neq0$. Finally, by using the explicit formula for the deformed bi-Hamiltonian structure given in the Appendix in Ref. 10, we obtain the unique bi-Hamilton structure of Equation (1.2). This completes the proof of Theorem 1.1.

\section{Data Availability}
No data was used for the research described in the article.


\begin{thebibliography}{99}
\bibitem{ref1}Camassa R, Holm D D. An integrable shallow water equation with peaked solitons. Phys Rev Lett. 1993, 71:1661-1664.
\bibitem{ref2}Camassa R, Holm D D, Hyman J M. A new integrable shallow water equation. Adv Appl Mech. 1994, 31: 1-33.
\bibitem{ref3}Anco S C, Recio E, Gandarias M L and Bruz M S. A nonlinear generalization of the Camassa-Holm equation with peakon solutions. Dynamical Systems, Differential Equations and Applications AIMS Proceeding. 2015, pp 29C37.
\bibitem{ref4}Fuchssteiner B. Some tricks from the symmetry-toolbox for nonlinear equations: generalizations of the Camassa-Holm equation. Phys D. 1996, 95 229-43
\bibitem{ref5}Anco S C,  Recio E. A general family of multi-peakon equations and their properties. J. Phys. A Math. Theor. 2019, 52, 125203.
\bibitem{ref6}Ablowitz M J, Segur H. Solitons and the inverse scattering transform. SIAM Stud Appl Math 1st Ed. Philadelphia: SIAM. 1981.
\bibitem{ref7}Ablowitz M J, Clarkson P A. Solitons, Nonlinear Evolution Equations and Inverse Scattering. Cambridge University Press. 1991.
\bibitem{ref8}Mikhailov AV, Sokolov V V. Symmetries of differential equations and the problem of integrability. In: Integrability, ed. by Mikhailov AV, Lecture Notes in Phys. Berlin: Springer. 2009, 767:19-88.
\bibitem{ref9}Olver P J. Applications of Lie Groups to Differential Equations. Springer, 2nd editio.; 1993.
\bibitem{ref10}Dubrovin B. On Hamiltonian perturbations of hyperbolic systems of conservation laws, II: universality of critical behaviour. Commun Math Phys. 2006, 267:117-139.
\bibitem{ref11}Dubrovin B, Liu SQ, Zhang Y. On Hamiltonian perturbations of hyperbolic systems of conservation laws I: quasi-triviality of bi-Hamiltonian perturbations. Comm Pure Appl Math. 2006, LIX:0559-0615.
\bibitem{ref12}da Silva P L, Freire I L. Integrability, existence of global solutions, and wave breaking criteria for a generalization of the Camassa-Holm equation.
Stud Appl Math. 2020,1C25
\bibitem{ref13}Baikov V A, Gazizov R, Ibragimov N. Approximate symmetries. Math USSR Sb. 1989, 64:427-441.
\bibitem{ref14}Freire I L. Conservation laws for self-adjoint first order evolution equation. J Nonl Math Phys. 2011, 18:279-290.
\bibitem{ref15}Freire I L. New conservation laws for inviscid Burgers equation. Comp Appl Math. 2012, 31:559-567.
\end{thebibliography}
\end{document}